\documentclass[twocolumn,showpacs,aps,prd,superscriptaddress,floatfix]{revtex4}

\usepackage{graphicx}
\usepackage{dcolumn}
\usepackage{bm}

\input babarsym.tex

\def\Vub {\ensuremath{V_{ub}}}

\def\btn {\ensuremath{B^{+} \to \tau^{+} \nu}\xspace}
\def\bmn {\ensuremath{B^{+} \to \mu^{+} \nu}\xspace}
\def\ben {\ensuremath{B^{+} \to e^{+} \nu}\xspace}
\def\btodx {\ensuremath{\Bub \to D^{(*)0} X^-}}

\def\eextra {\ensuremath{E_{\mathrm{extra}}}\xspace}

\def\nulb  {\ensuremath{\bar{\nu}_{\ell}}}
\def\nueb  {\ensuremath{\bar{\nu}_{e}}}

\def\nutb  {\ensuremath{\bar{\nu}_{\tau}}}

\def\tautoenunu {\ensuremath {\tau^+ \to e^+ \nu \nub}}

\def\tautomununu {\ensuremath {\tau^+ \to \mu^+ \nu \nub}}

\def\tautopinu {\ensuremath {\tau^+ \to \pi^+ \nub}}

\def\tautopipiznu {\ensuremath {\tau^+ \to \pi^+ \pi^{0} \nub}}
\def\tautopipiz {\ensuremath {\tau^+ \to \pi^+ \pi^{0} \nub}}

\def\bcount    {\ensuremath {383 \times 10^{6}} }
\def\bcount    {\ensuremath {383 \times 10^{6}} }
\def\onlumi    {\ensuremath { 346  \invfb\ }}
\def\offlumi   {\ensuremath { 36.3 \invfb\  }}

\def\nBB   {\ensuremath {383 \times 10^{6}} }

\def\pipiz {\ensuremath { \pi^+ \pi^{0} }}

\newcommand{\BABARPubYear}    {07}
\newcommand{\BABARPubNumber}  {046}

\newcommand{\SLACPubNumber} {12736}

\begin{document}
\noindent\babar-PUB-\BABARPubYear/\BABARPubNumber \\
SLAC-PUB-\SLACPubNumber \\

\title{A Search for \boldmath{\btn} decays with hadronic \textit{B} tags}


%
\author{B.~Aubert}
\author{M.~Bona}
\author{D.~Boutigny}
\author{Y.~Karyotakis}
\author{J.~P.~Lees}
\author{V.~Poireau}
\author{X.~Prudent}
\author{V.~Tisserand}
\author{A.~Zghiche}
\affiliation{Laboratoire de Physique des Particules, IN2P3/CNRS et Universit\'e de Savoie, F-74941 Annecy-Le-Vieux, France }
\author{J.~Garra~Tico}
\author{E.~Grauges}
\affiliation{Universitat de Barcelona, Facultat de Fisica, Departament ECM, E-08028 Barcelona, Spain }
% ...
\collaboration{The \babar\ Collaboration}
\noaffiliation

\date{\today}

\begin{abstract}
\noindent We present a search for the decay \btn\ using \bcount $\B\Bbar$ pairs   
collected at the $\Y4S$ resonance with the \babar\ detector 
at the SLAC PEP-II $B$ Factory. 
We select a sample of events with one completely reconstructed tag $B$ in a hadronic
decay mode (\btodx), and
examine the rest of the event to search for a $\btn$ decay. 
We identify the $\tau$ lepton  in the following modes: $\tautoenunu$, $\tautomununu$,
$\tautopinu$ and $\tautopipiznu$. 
We find a 2.2 $\sigma$ 
excess in data and measure a branching fraction of 
$\mathcal{B}(\btn)=( 1.8^{+0.9}_{-0.8}(\mbox{stat.})\pm 0.4(\mbox{bkg. syst.}) \pm 0.2 (\mbox{other syst.})) \times 10^{-4}$. 
We calculate the product of the $B$ meson decay constant $f_{B}$ and $|\Vub|$ to be
$f_{B}\cdot|\Vub| = (10.1^{+2.3}_{-2.5}(\text{stat.})^{+1.2}_{-1.5}(\text{syst.}))\times10^{-4}$~GeV
\end{abstract}

\maketitle
The study of the purely leptonic decay $\btn$~\citep*{cc} is 
of particular interest because it is sensitive to the product of the $B$ meson decay constant $f_{B}$, 
and the absolute value of Cabibbo-Kobayashi-Maskawa matrix element \mbox{$\Vub$~\citep*{c,km}}. 
In the Standard Model (SM), the decay 
proceeds via quark annihilation into a $W^{+}$ boson, with a
branching fraction given by:
\begin{equation}
\label{eqn:br}
\mathcal{B}(B^{+} \rightarrow {\taup} \nu)= 
\frac{G_{F}^{2} m^{}_{B}  m_{\tau}^{2}}{8\pi}
\left[1 - \frac{m_{\tau}^{2}}{m_{B}^{2}}\right]^{2} 
\tau_{\Bu} f_{B}^{2} |\Vub|^{2},
\end{equation}
where  
$G_F$ is the Fermi constant,
$\tau_{\Bu}$ is the $\Bu$ lifetime, and
$m^{}_{B}$ and $m_{\tau}$ are the $\Bu$ meson and $\tau$ lepton masses.
Using $|\Vub| = (4.31 \pm 0.30)\times 10^{-3}$ from experimental measurements of semileptonic B decays~\cite{pdg2004} and 
$f_{B} = 0.216 \pm 0.022$ GeV from lattice QCD~\citep*{fb},
the SM estimate of the branching fraction for $\btn$ is $(1.5 \pm 0.4)\times 10^{-4}$.

The process \btn is also  sensitive to extensions of the SM. 
For instance, in two-Higgs doublet models~\cite{higgs} and in the MSSM~\cite{Isidori2006pk,Akeroyd:2003zr}
 it could be mediated by charged Higgs bosons.
The branching fraction measurement can therefore also be used to constrain
the parameter space of extensions to the SM.

The \bmn\ and \ben\ decays are significantly helicity suppressed with respect 
to the \btn channel. However, a search for $\btn$ is experimentally 
more challenging, due to the presence of multiple neutrinos in the
final state, which makes the experimental signature less distinctive.
In a previously published analysis using a sample of 
$383 \times 10^6$ $\FourS\to\B\Bbar$ decays, based on the reconstruction of a semileptonic \textit{B} decay on the tag side, 
the \babar\ collaboration set an upper limit 
$\mathcal{B}(\btn)<1.7 \times 10^{-4} \, 
\textrm{ at the 90\% confidence level (CL)}$~\citep{taunusemilep}.
The Belle Collaboration has reported evidence from a search for this decay and the branching fraction was measured to be
$\mathcal{B}(\btn) = (1.79^{+0.56}_{-0.49}(\mbox{stat.}) ^{+0.46}_{-0.51}(\mbox{syst.})) \times 10^{-4}$
~\citep{belle}.

The data used in this analysis were collected with the \babar\ detector
at the \pep2\ storage ring. 
The sample corresponds to an integrated
luminosity of \onlumi at the \FourS\ resonance (on-resonance) 
and \offlumi taken at $40\mev$  below the \FourS\ resonance (off-resonance). 
The on-resonance sample contains $\nBB$  $B\bar{B}$ decays. 
The detector is described in detail elsewhere~\citep{babar}. 
Charged-particle trajectories are measured in the tracking system
composed of a five-layer silicon vertex detector and a 40-layer drift chamber (DCH),
operating in a  1.5~T solenoidal magnetic field.
A Cherenkov detector is used for $\pi$--$K$ discrimination, a CsI calorimeter
for photon detection and electron identification, and 
the flux return of the solenoid, which consists of layers of steel
interspersed with resistive plate chambers or limited streamer tubes, for muon
and neutral hadron identification.

In order to estimate signal selection efficiencies and to study physics backgrounds,
we use a \babar\ Monte Carlo (MC) simulation based on GEANT4 \cite{geant4}.
In MC simulated signal events one $B^+$ meson decays to
$\tau^+\nu$ and the other into any final state. 
The \BB\ and continuum MC samples are, respectively, equivalent to approximatively three times and
1.5 times the accumulated data sample.
Beam-related background and detector noise are taken from data 
and overlaid on the simulated events.

We reconstruct an exclusive decay of one of the $B$ mesons in the event (tag $B$)
and examine the remaining particle(s) for the
experimental signature of \btn. 
In order to avoid experimenter bias, the 
signal region in data is  blinded until the final yield
extraction is performed.

The tag $B$ candidate is reconstructed in the set of hadronic $B$ decay modes 
\btodx~\cite{cc}, where $X^-$ denotes a system of
charged and neutral hadrons with total charge $-1$
composed of $n_1\pi^{\pm}$, $n_2K^{\pm}$, $n_3\KS$,  $n_4\piz$, where $n_1 + n_2 \leq
5$,  $n_3  \leq  2$,  and  $n_4  \leq  2$.
We  reconstruct $\Dstarz \ra \Dz\piz, \Dz\gamma$; 
$\Dz\ra K^-\pi^+$, $K^-\pi^+\piz$, $K^-\pi^+\pi^-\pi^+$,  $\KS\pi^+\pi^-$ and  $\KS \ra \pi^+\pi^-$. 
The kinematic consistency of tag $B$ candidates 
is checked with
the beam energy-substituted mass $\mes = \sqrt{s/4 -
\vec{p}^{\,2}_B}$ and the energy difference 
$\Delta E = E_B - \sqrt{s}/2$. Here $\sqrt{s}$ is the total
energy in the \FourS center-of-mass (CM) frame, and $\vec{p}_B$ and $E_B$
denote, respectively, the momentum and energy of the tag $B$ candidate in the CM
frame.  The resolution on $\Delta E$ is measured to be $\sigma_{\Delta E}=10-35\mev$, depending on
the decay mode; we require $|\Delta E|<3\sigma_{\Delta E}$.
The purity ${\cal P}$ 
of each reconstructed $B$ decay mode
is estimated, using on-resonance data, 
as the ratio of the number of peaking events with \mes$>
5.27$\gevcc to the total number of events in the same range.
If multiple tag $B$ candidates are reconstructed, the one with the highest purity ${\cal P}$ is selected.
If more than one candidate with the same purity is reconstructed, the one with the lowest value of
$|\Delta E|$ is selected. 
From the dataset obtained as described above, we consider only those
events in which 
the tag $B$ is reconstructed in the decay modes of highest purity ${\cal P}$.
The set of decay modes used is defined by the requirement that the purity
of the resulting sample is not less than 30\%.

The background consists of $\epem\ra q\bar q \ (q=u,d,s,c)$ events and
other $\FourS\to\BzBzb$ or \BpBm decays
in which the tag $B$ candidate is mis-reconstructed using particles 
coming from both $B$ mesons in the event.
To  reduce the $\epem\ra q\bar q$ background,  we require 
$|\cos{\theta_{TB}^*}|<0.9$, where $\theta_{TB}^*$ is 
the angle in the CM frame between the thrust
axis~\cite{thrust} of the tag $B$ candidate and the thrust axis of the 
remaining reconstructed charged and neutral candidates.

\begin{figure}[!thb]
  \begin{center}
    \includegraphics[width=0.90\linewidth]{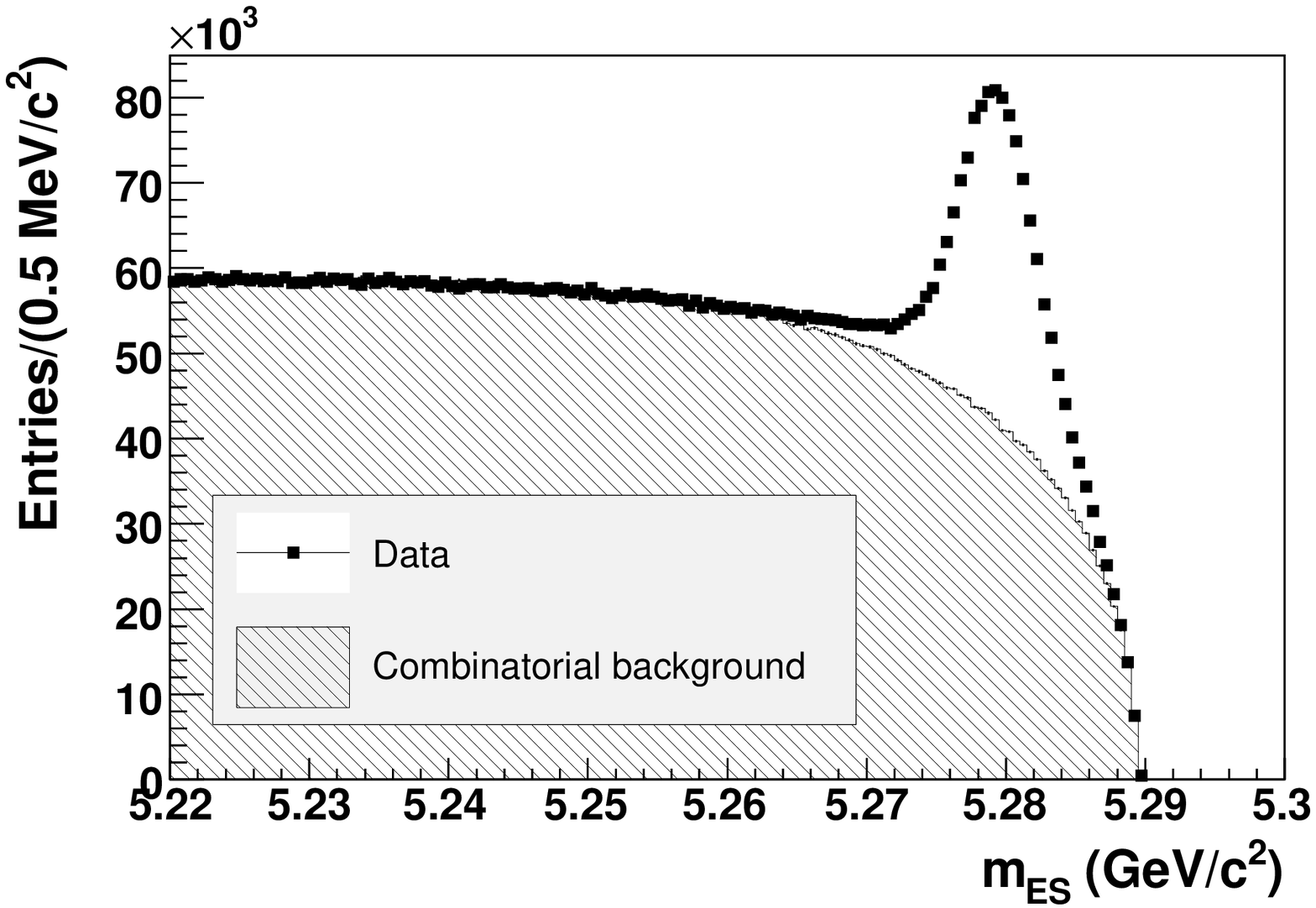}

    \caption{Distribution of the energy substituted mass, \mes, 
      of the tag $B$ candidates in data. The combinatorial background is overlaid. 
    } 
    \label{fig:mesfit}
  \end{center}
\end{figure}

In order to determine the number of correctly reconstructed $\Bu$ decays,
we classify the background events in four  categories: $\epem\ra c\bar c$;
$\epem\ra\uubar,\ddbar,\ssbar$; \FourS\ra\BzBzb; and \FourS\ra\BpBm. 
The \mes shapes of these background distributions are taken from MC simulation.
The normalization of the  $\epem\ra c\bar c$ and $\epem\ra \uubar,\;\ddbar,\;\ssbar$ backgrounds is taken from 
off-resonance data, scaled by the luminosity and corrected for the 
different selection efficiencies evaluated with the MC.
The normalization of the \BzBzb, \BpBm components are 
obtained by means of  a $\chi^2$ fit to the \mes distribution in 
the data sideband region ($5.22\gevcc<\mes<5.26\gevcc$). 
The number of background events in the signal region ($\mes>5.27\gevcc$) is extrapolated from the fit and 
subtracted from the data. We estimate the total number of tagged $B$'s in the data to be
$N_{B} = (5.92 \pm 0.11\textrm{(stat)}) \times 10^{5}$.
Figure \ref{fig:mesfit} shows the tag $B$ candidate \mes\ distribution,
with the combinatorial background, 
estimated as the sum of the four components described above, overlaid.  

After the reconstruction of the tag $B$ meson, a set of selection criteria 
is applied to the rest of the event (recoil) in order to 
enhance the sensitivity to \btn decays.
We require the presence of only one well-reconstructed charged track (signal track) 
with charge opposite to that of the tag $B$. The signal track is required to have at least
12 hits in the DCH, momentum transverse to the 
beam axis, $p_{T}$, greater than 0.1$\gevc$, and
the point of closest approach to the interaction point
less than 10\cm along the beam axis and less than 1.5\cm transverse 
to it.

The $\tau$ lepton is identified in four decay modes
constituting approximately 71\% of the total $\tau$ decay width: $\tautoenunu$, $\tautomununu$, $\tautopinu$, and \tautopipiznu .
Particle identification criteria on the signal track are used to separate the four 
categories. 
The  \tautopipiznu\ sample is obtained by associating the signal track, identified 
as pion, with a $\pi^0$ reconstructed from a pair  of neutral clusters with invariant mass 
between 0.115 and 0.155 \gevcc and total energy greater than 250 MeV. 
In case of multiple  \pipiz\ candidates, the one with largest
center-of-mass momentum $p^*_{\pi^+\piz}$ is chosen.

We place a mode-dependent cut on $|\cos\theta^*_{TB}|$ to reduce the background 
due to continuum events and incorrectly reconstructed tag $B$ candidates (combinatorial).
The remaining sources of background consists
of $\BpBm$ events in which the tag $B$ meson was correctly reconstructed and the
recoil contains one track and additional particles that are not reconstructed by the tracking detectors and
calorimeter. 
MC simulation shows that most of this background is from semileptonic $B$ decays. 

We define the discriminating variable \eextra as the sum of the energies 
of the neutral clusters not associated
with the tag $B$ or with the signal $\pi^0$ from the \tautopipiz\ mode,
and passing a minimum energy requirement. The required energy depends 
on the selected signal mode and on the calorimeter region involved and varies from
50 to 70 \mev.
Signal events tend to peak at low \eextra values, whereas background events, which contain additional sources 
of neutral clusters, are distributed toward higher \eextra values.

Other variables used to discriminate between signal and  background are the CM
momentum of the signal candidates, the multiplicities of low  $p_T$ 
charged tracks and of $\pi^0$ candidates in the recoil, and 
the direction of the missing momentum four-vector in the CM frame. 
For the  $\tautopipiznu$ mode, we exploit the presence of the $\pi^0$ in the 
final state and the dominance of the decay through the $\rho^+$ resonance 
by means of the combined quantity 
$x_{\rho} = [(m_{\pi^+\pi^0}-m_\rho)/(\Gamma_\rho)]^2 
+ [(m_{\gamma\gamma}-m_{\pi^0})/(\sigma_{\pi^0})]^2$,
where $m_{\pi^+\pi^0}$ is the reconstructed invariant mass of the \pipiz\ candidate, $m_{\gamma\gamma}$ 
is  the reconstructed invariant mass of the $\pi^0$ candidate, $m_\rho$ and $\Gamma_\rho$ are the 
nominal values~\cite{pdg2004} for the $\rho$ mass and width, $m_{\pi^0}$ is the nominal $\pi^0$ mass and 
$\sigma_{\pi^0} = 8 \mevcc$ 
is the experimental resolution on the $\pi^0$ mass determined from data. 

We optimize the selection by maximizing $s/\sqrt{s+b}$ using the \BpBm
MC  and signal MC, where $b$ is the expected background from \BpBm events 
and $s$ is the expected number of signal events 
in the hypothesis of a branching fraction 
of $1 \times 10^{-4}$. The optimization is performed separately for each $\tau$ decay mode and 
with all the cuts applied simultaneously in order to take into account any correlations  
among  the discriminating variables.
The optimized signal selection cuts are reported in Table~\ref{tab:selcuts}.
\begin{table}[!tbhp]
\caption{Optimized selection criteria  for each $\tau$ decay mode. }
\begin{center}
\begin{tabular}{lrrrr} \hline \hline 
      Variable                   & $e^+$      & $\mu^+$    & $\pi^+$     & $\pi^+ \pi^{0}$ \\
      \hline 
      \eextra (\gev)             &  $< 0.160$ & $< 0.100$  & $< 0.230$   & $< 0.290$       \\ 
      \piz multiplicity          &     0      &        0   &    $\leq 2$ &  --             \\
      Track multiplicity         & 1          &  1         &    $\leq 2$ &  1              \\
      $|\cos\theta^*_{TB}|$      & $\leq 0.9$ & $\leq 0.9$ & $\leq 0.7$  & $\leq 0.7$      \\ 
      $p^{*}_{\rm{trk}} (\gevc)$ & $<1.25$    &  $<1.85$   & $>1.5$      &  --             \\
      $\cos \theta^*_{\rm{miss}}$& $<0.9$     &  --        & $<0.5$      & $<0.55$         \\
      $p^{*}_{\pipiz} (\gevc) $  & --         &  --        & --          & $>1.5$          \\
      $ x_\rho $                 & --         &  --        & --          & $<2.0$          \\
      $ E_{\piz} $~(\gev)        & --         &  --        & --          & $>0.250$        \\
      \hline \hline
\end{tabular}
\end{center}
\label{tab:selcuts}
\end{table}

We compute the signal selection efficiency as the ratio of the number of 
signal MC events passing the selection criteria to the number of signal events
that have a correctly reconstructed tag \B\ candidate in the signal region $\mes > 5.27 \gevcc$. 
We evaluate the efficiencies on a signal MC sample which is distinct 
from the sample used in the optimization procedure. 
A small cross-feed in some modes is estimated from MC and is taken 
into account in the computation of the total efficiency.

The total efficiency for each selection is given by:

\begin{equation}
  \varepsilon_i	= \sum_{j=1}^{n_{\text{dec}}} \varepsilon_i^j f_j \,\, ,
\label{eq:eff}
\end{equation}
where $\varepsilon_i^j$ is the efficiency of the selection $i$
for the $\tau$ decay mode $j$, $n_{\text{dec}}=7$ is the number of $\tau$
decay modes that can contribute to the reconstructed modes and $f_j $ are the 
fractions of the $\tau$ decay mode as estimated from the signal MC sample with a reconstructed tag $B$. 
Table~\ref{tab:eff} shows the estimated efficiencies.

\begin{table}[!tbhp]
  \caption{ 
Efficiency (in percent) of the most relevant $\tau$ decay modes (rows)
to be selected in one of the four modes considered in this analysis (column).
The  All decay row shows the selection efficiency of each reconstruction mode,
adding the contribution from the previous rows, weighted by the decay abundance at
the tag selection level $f_j$. The last row shows the total signal selection efficiency.
The uncertainties are statistical only.
}
    \begin{center} 
    \begin{tabular}{lcccc} \hline \hline
 Mode  & $e^+$ & $\mu^+$ & $\pi^+$ & $\pi^+ \pi^{0}$ \\
\hline 
$e^+$ & {\bf 19.3 $\pm$ 1.1}  &  0 &  0.4 $\pm$ 0.2 &  0 \\
$\mu^+$ &  0 & {\bf 10.8 $\pm$ 0.9}  &  1.3 $\pm$ 0.3 &  0 \\
$\pi^+$ &  0 &  0.1 $\pm$ 0.1 & {\bf 19.7 $\pm$ 1.3}  &  0.5 $\pm$ 0.2\\
$\pi^+ \pi^{0}$  &  0 &  0  &  1.5 $\pm$ 0.2 & {\bf  7.0 $\pm$ 0.5} \\
$\pi^+ \pi^+ \pi^-$   &  0 &  0 &  0 &  0 \\
$\pi^+ \pi^0 \pi^0$ &  0 &  0 &  0.2 $\pm$ 0.1 &  1.8 $\pm$ 0.4\\
Other &  0 &  0 &  0.3 $\pm$ 0.2 &  0.1 $\pm$ 0.1\\
\hline \hline
 All dec. $\epsilon_i$:  & {\bf  3.1$\pm$0.2}  & {\bf  1.7$\pm$0.1}  & {\bf  2.9$\pm$0.2}  & {\bf  2.2$\pm$0.2} \\ \hline
 Total: &\multicolumn{4}{c}{\bf 9.8 $\pm$ 0.3}\\ \hline
\end{tabular}
  \end{center}
  \label{tab:eff}
\end{table}

To determine the  expected number of background events in the data, 
we use the final selected data samples with $\eextra$ between  0 and 2.4 $\gev$. 
We first perform an extended unbinned maximum likelihood fit to
the \mes distribution
in the  $\eextra$ sideband  region $ 0.4 \gev< \eextra < 2.4 \gev$
of the final sample.
For the peaking component of the background we use a probability density function (PDF) 
which is  a Gaussian function joined
to an exponential tail (Crystal Ball function)~\cite{crystalball}. As a PDF for the non-peaking component,
we use a phase space motivated threshold function (ARGUS function)~\cite{arguspdf}.
From this fit, we determine a peaking yield  $N_{pk}^{\rm{side},\rm{data}}$ 
and signal shape parameters, to be used in later fits. 
We apply the same procedure to \BpBm MC events which 
pass the final selection and determine the peaking yield $N_{pk}^{\rm{side},\rm{MC}}$.
To determine the MC peaking yield in the \eextra signal region 
$N_{pk}^{\rm{sig},\rm{MC}}$, 
we fit \mes in the \eextra signal region of the \BpBm MC sample
with the Crystal Ball parameters fixed to the values determined
in the \eextra sideband fits described above.
Analogously, we fit the  \mes distribution of data in the \eextra signal region 
to extract the combinatorial  background $n_{\text{comb}}$,
evaluated as the integral of the ARGUS shaped component 
in the $\mes > 5.27 \gevcc$ region.
We estimate the total expected background  in the signal region as:
\begin{equation}
b = \frac{ N_{pk}^{\rm{sig},\rm{MC}}} {N_{pk}^{\rm{side},\rm{MC}}} \times N_{pk}^{\rm{side},\rm{data}} + n_{\text{comb}}.
\label{eq:bgformula}
\end{equation}

\begin{figure}[htb]
\includegraphics[width=0.9\linewidth]{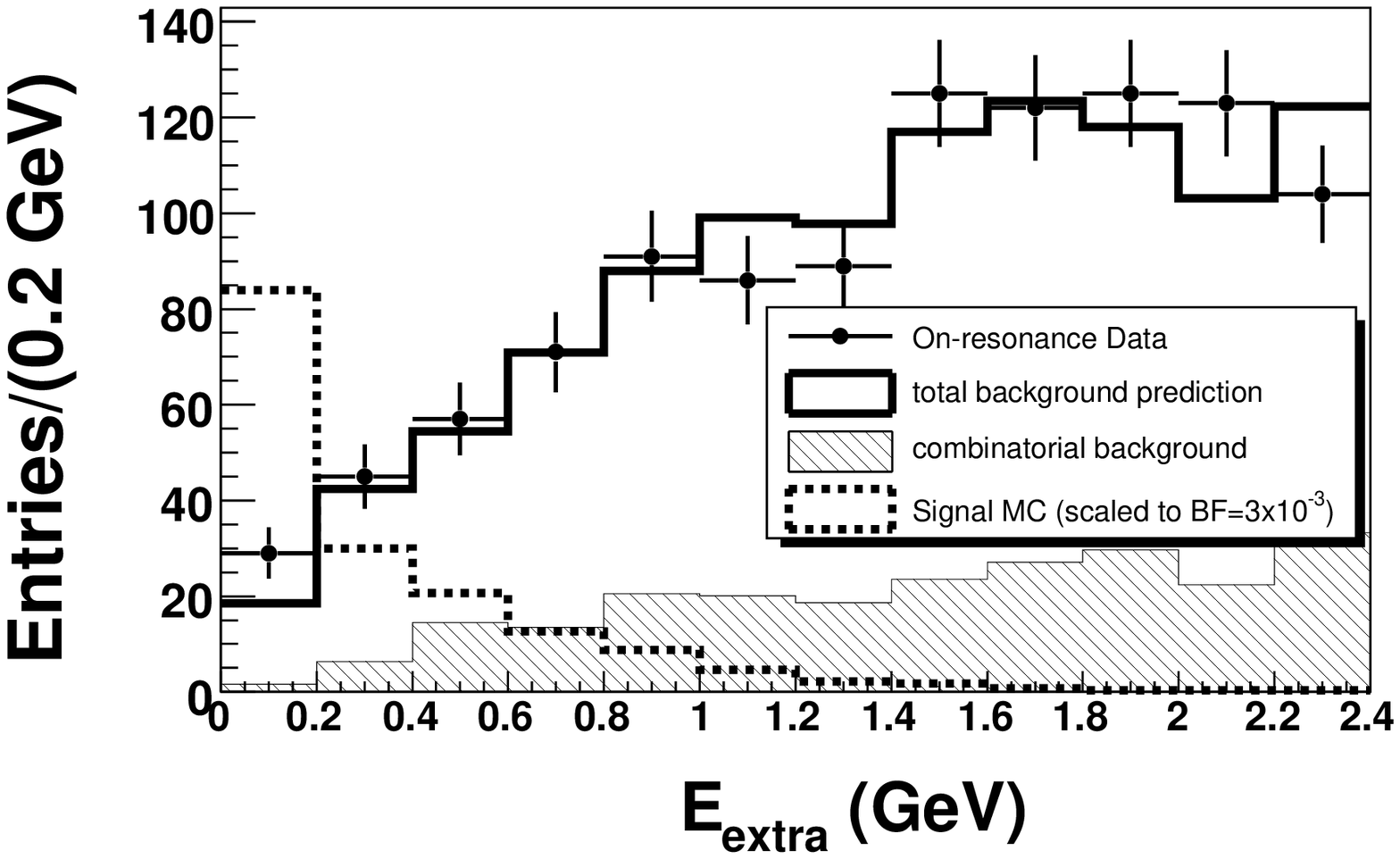} \\
\includegraphics[width=0.9\linewidth]{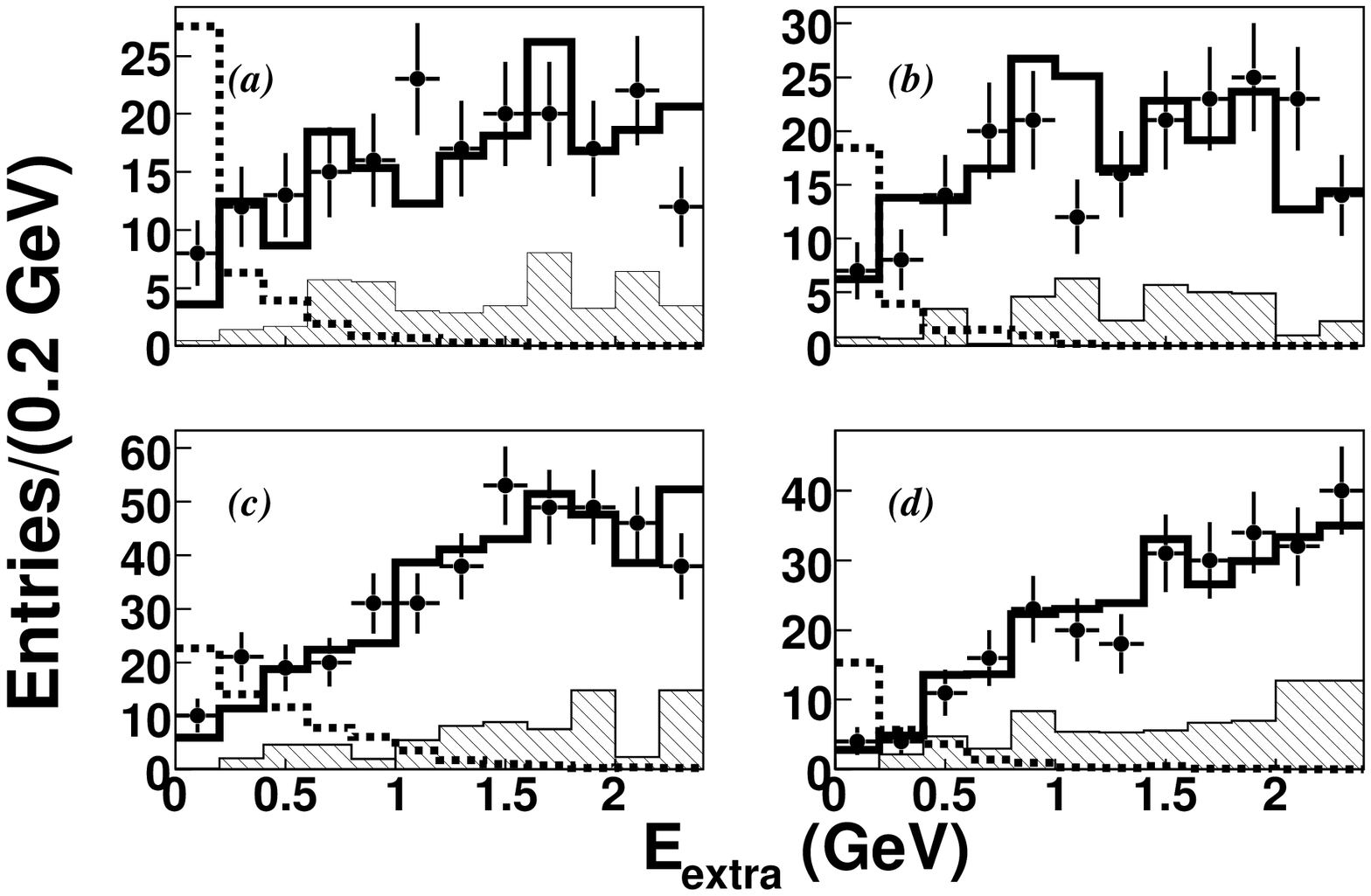} \\
\caption{$\eextra$ distribution after all selection criteria 
have been applied. The upper plot shows the distribution of all the modes combined while lower plots show  
the (a) $\tautoenunu$, (b) $\tautomununu$, (c) $\tautopinu$, and (d) $\tautopipiznu$
modes separately. The on-resonance data (black dots) distribution is compared with the total background prediction
(continuous histogram). The hatched histrogram represents the combinatorial background component.
$B^+\to\tau^+\nu$ signal MC (dashed histogram), normalized to a branching fraction of $3\times10^{-3}$ for illustrative purposes,
 is shown for comparison. }
\label{fig:eextra}
\end{figure}

After finalizing the signal selection criteria, we measure the
yield of events in each decay mode in on-resonance data. 
Table~\ref{tab:unblres} reports the
number of observed events together with the expected number of 
background events, for each $\tau$ decay mode.
Figure \ref{fig:eextra} shows the \eextra distribution for data and
expected background at the end of the selection. The signal MC,
normalized to a branching fraction of $3\times10^{-3}$ for illustrative purposes, 
is overlaid for comparison.
The \eextra distribution is also plotted separately for each $\tau$ decay mode.  

\begin{table}
\caption{
Observed number of on-resonance data events in the signal region compared
 with the number of expected background events.}
\begin{center}
\begin{tabular}{lcc} \hline 
$\tau$ decay mode   & Expected background  &  Observed   \\ 
\hline \hline
\tautoenunu	  & 1.47  $\pm$ 1.37   & 4  \\ 
\tautomununu      & 1.78  $\pm$ 0.97   & 5  \\ 
\tautopinu        & 6.79  $\pm$ 2.11   & 10 \\ 
\tautopipiznu     & 4.23  $\pm$ 1.39   & 5  \\ 
\hline
All modes    & 14.27 $\pm$  3.03 & 24  \\ \hline \hline
\end{tabular}
\end{center}
\label{tab:unblres}
\end{table}

We combine the results on the observed number of events  $n_i$ and on the expected background
$b_i$  from each of the four signal decay modes ($n_{ch}$) using the estimator
\mbox{$Q = {\cal L}(s+b)/{\cal L}(b)$},
where ${\cal L}(s+b)$ and ${\cal L}(b)$ are the
likelihood functions for signal plus background and background-only
hypotheses, respectively:
\begin{equation}
  {\cal L}(s+b) \equiv
  \prod_{i=1}^{n_{ch}}\frac{e^{-(s_i+b_i)}(s_i+b_i)^{n_i}}{n_i!},
        \;
  {\cal L}(b)   \equiv
  \prod_{i=1}^{n_{ch}}\frac{e^{-b_i}b_i^{n_i}}{n_i!}.
  \label{eq:lb}
\end{equation}
The  estimated number of signal candidates $s_i$ in data, for each decay mode, is related to the 
\btn branching fraction by:
\begin{equation}
s_i = \frac {   \varepsilon^{\rm{tag}}_{\rm{sig}} } { \varepsilon^{\rm{tag}}_{B} }
      N^{\rm{tag}}_{\Bu}\varepsilon_i  \mathcal{B}(\btn),
\end{equation}
where $N^{\rm{tag}}_{\Bu}$ is the number of tag $\Bu$ mesons 
correctly reconstructed, 
$ \varepsilon^{\rm{tag}}_{B}$ and $\varepsilon^{\rm{tag}}_{\rm{sig}}$ are the tag $B$ efficiencies in 
generic $B\bar{B}$ and signal events respectively, 
and $\varepsilon_i$ are the signal efficiencies defined in equation \ref{eq:eff}.
We fix the ratio
$\varepsilon^{\rm{tag}}_{\rm{sig}} / \varepsilon^{\rm{tag}}_{B}  = 0.939\pm0.007\textrm{(stat.)}$
to the value obtained from MC simulation.

We estimate the branching fraction
(including statistical uncertainty and uncertainty from the background) 
by scanning over signal branching fraction hypotheses 
and computing the value of $\mathcal{L}(s+b)/\mathcal{L}(b)$ for each
hypothesis. The branching fraction is the hypothesis which minimizes the likelihood ratio 
$-2 \ln \textrm{Q}= -2 \ln(\mathcal{L}(s+b)/\mathcal{L}(b))$,
and we determine the  statistical uncertainty  by finding the points on the likelihood scan that
occur at one unit above the minimum. 

The dominant uncertainty on the background predictions $b_i$ is
due to the finite \BpBm MC statistics. 
We also check possible systematic effects in the 
estimation of combinatorial background by means of a sample
of events with looser selection requirements; we
find it to be negligible with respect to the statistical
uncertainty. The background uncertainty is incorporated in the likelihood
definition used to extract the branching fraction, by
convolving it with a Gaussian function with
standard deviation equal to the error on $b_i$~\cite{giunti}.

The other sources of systematic uncertainty in the determination of the $\btn$
branching fraction come from the estimation of the tag yield and efficiency
and the reconstruction efficiency of the signal modes.
We estimate the systematic uncertainty 
on the tag $B$ yield and reconstruction efficiency by varying the MC 
\BpBm non-peaking component of the \mes shape,
assigning a systematic uncertainty of 3\% on the branching fraction.
The systematic uncertainties due to mismodeling of charged
particle tracking efficiency, \eextra shape, particle identification 
efficiency, $\pi^0$ reconstruction and signal MC statistics 
depend on the $\tau$ decay mode.
The uncertainty on the branching fraction is evaluated for each mode
separately. We obtain the total contributions due to  tracking and \eextra systematics
by adding linearly the contributions of each decay channel.
The total contributions due to MC statistics and particle identification
are obtained by adding systematics uncertainties of each reconstruction mode in quadrature. 

We check the low $p_T$ charged track multiplicity distribution agreement 
between data and MC with a sample enriched in background by loosening the 
selection criteria. The disagreement, which is mode dependent, 
is quantified by comparing the MC PDF with the data PDF. 
We correct the MC to reproduce the distribution in data and apply
the correction to the signal MC distribution. We take 100\% of the correction
as a systematic uncertainty, resulting in a total 
systematic uncertainty of 5.8\% on the branching fraction.

The systematic uncertainty due to the \eextra mismodeling is determined
by means of a data sample containing events with two 
non-overlapping tag $B$ candidates.
The sample is selected by reconstructing a second $B$ meson in a
hadronic decay mode \btodx on the recoil of the tag $B$.
In addition to the requirements on the tag $B$ described above, we 
consider only second $B$ candidates satisfying $|\DeltaE|<50 \mev$ and
$\mes>5.27\gevcc$ having opposite charge to that of the tag $B$. 
If multiple candidates are reconstructed, the one with the
highest purity ${\cal P}$ is selected.
We compare the distribution of the total energy of the unassigned neutral 
clusters \eextra in data and in MC.
We compute the ratio of the number of events in the signal region of each 
$\tau$ mode to the total number of events in the
sample. For each $\tau$ mode, we evaluate the systematic uncertainty,
comparing the ratio estimated from MC to the ratio estimated 
from data. This procedure results in a 8.8\% systematic uncertainty
on the branching fraction. 
Table \ref{tab:SignalEffSys} shows the 
contributions in percent to the systematic uncertainties on the branching 
fraction.

\begin{table}
\caption{Contributions (in percent) to the systematic uncertainty on the 
branching fraction due to signal selection efficiency for different 
selection modes.}
\begin{center}
\begin{tabular}{lccccc} \hline \hline
Source of systematics    & $e^+$ & $\mu^+$ & $\pi^+$ & $\pi^+ \pi^{0}$ & Total \\
\hline 
MC statistics             & 3.1 & 0.6  & 1.5 & 2.6 & 4.3 \\
Particle Identification   & 1.5 & 1.3  & 0.2  & 0.2 & 2.0 \\
$\piz$                    & --  & --   & --   & 1.4 & 1.4 \\
\hline
Tracking                  & 3.7 & 0.4  & 0.1  & 1.6 & 5.8 \\
\eextra			  & 4.7 & 0.6  & 0.9  & 2.6 & 8.8 \\
\hline
Signal  $B$               &     &      &      &     & 11.6 \\
Tag $B$                   &     &      &      &     &   3 \\
\hline
Total                     &     &      &      &     &   12 \\
\hline \hline
\end{tabular}		    
\end{center}
\label{tab:SignalEffSys}
\end{table}

In summary, we measure the branching fraction 
\begin{equation}
\mathcal{B}(\btn)=(1.8^{+0.9}_{-0.8}\pm 0.4 \pm 0.2) \times 10^{-4},
\label{eqn:bf}
\end{equation}
where the first error is statistical, the second is due to the background uncertainty, and the third is due 
to other systematic sources.
Taking into account the uncertainty on the expected background, as described above,  we obtain a significance 
of 2.2~$\sigma$.

Using Eq. \ref{eqn:br}, we calculate the product 
 of the $B$ meson decay constant $f_{B}$ and $|\Vub|$ to be
\mbox{$f_{B}\cdot|\Vub| = (10.1^{+2.3}_{-2.5}(\text{stat.})^{+1.2}_{-1.5}(\text{syst.}))\times10^{-4}$~GeV}.
We also measure the 90\% C.L. upper limit using the $CL_s$ method ~\cite{CLs} 
to be $ \mathcal{B}(\btn)  < 3.4 \times 10^{-4}$.

The combination of this measurement with the \babar\ result obtained using semileptonic tags, 
based on a statistically 
independent data sample, and reported in ~\cite{taunusemilep}, yields: 
\begin{equation}
\mathcal{B}(\btn) = (1.2 \pm 0.4_{\mbox{stat.}} \pm 0.3_{\mbox{bkg.}} \pm 0.2_{\mbox{syst.}} ) \times 10^{-4}.
\label{eqn:bfcombined}
\end{equation}
The significance of the combined result is  2.6 $\sigma$ including the uncertainty on the expected 
background~(3.2 $\sigma$ if this uncertainty is not included).

We are grateful for the excellent luminosity and machine conditions
provided by our \pep2\ colleagues, 
and for the substantial dedicated effort from
the computing organizations that support \babar.
The collaborating institutions wish to thank 
SLAC for its support and kind hospitality. 
This work is supported by
DOE
and NSF (USA),
NSERC (Canada),
CEA and
CNRS-IN2P3
(France),
BMBF and DFG
(Germany),
INFN (Italy),
FOM (The Netherlands),
NFR (Norway),
MIST (Russia),
MEC (Spain), and
STFC (United Kingdom). 
Individuals have received support from the
Marie Curie EIF (European Union) and
the A.~P.~Sloan Foundation.

\bibliography{paper}

\begin{thebibliography}{17}
\bibitem[{cc()}]{cc}
\bibinfo{note}{Charge-conjugate modes are implied throughout the paper.}
\end{thebibliography}

\end{document}